\documentclass[runningheads]{llncs}

\usepackage{graphicx}
\usepackage{amsmath}
\usepackage{amssymb}
\usepackage{multirow}
\usepackage{cite}
\usepackage{bm}
\usepackage{xcolor}

\begin{document}

\title{Intelligent image synthesis to attack a segmentation CNN using adversarial learning}

\author{Liang Chen\inst{1,2} \and
Paul Bentley\inst{2} \and
Kensaku Mori \inst{3} \and
Kazunari Misawa \inst{4} \and
Michitaka Fujiwara \inst{5} \and
Daniel Rueckert \inst{1}}

\institute{Department of Computing, Imperial College London, 180 Queen's Gate, London, SW7 2AZ, UK \\
\email{liang.chen12@imperial.ac.uk}\and
Department of Medicine, Imperial College London, Fulham Palace Road, London, W6 8RF, UK \and
Graduate School of Informatics, Nagoya University, Nagoya, 464-8603, Japan \and
The Aichi Cancer Center, Nagoya, 464-8681, Japan \and
Nagoya University Hospital, Nagoya, 466-8560, Japan
}

\maketitle

\begin{abstract}
Deep learning approaches based on convolutional neural networks (CNNs) have been successful in solving a number of problems in medical imaging, including image segmentation. In recent years, it has been shown that CNNs are vulnerable to attacks in which the input image is perturbed by relatively small amounts of noise so that the CNN is no longer able to perform a segmentation of the perturbed image with sufficient accuracy. Therefore, exploring methods on how to attack CNN-based models as well as how to defend models against attacks have become a popular topic as this also provides insights into the performance and generalization abilities of CNNs. However, most of the existing work assumes unrealistic attack models, i.e. the resulting attacks were specified in advance. In this paper, we propose a novel approach for generating adversarial examples to attack CNN-based segmentation models for medical images. Our approach has three key features: 1) The generated adversarial examples exhibit anatomical variations (in form of deformations) as well as appearance perturbations; 2) The adversarial examples attack segmentation models so that the Dice scores decrease by a pre-specified amount; 3) The attack is not required to be specified beforehand. We have evaluated our approach on CNN-based approaches for the multi-organ segmentation problem in 2D CT images. We show that the proposed approach can be used to attack different CNN-based segmentation models.

\end{abstract}

\section{Introduction}

CNNs have been amongst the most popular model for image classification and segmentation problems thanks to their efficiency and effectiveness in learning representative image features. However, it has been widely reported that even the most well-established CNN models such as the GoogLeNet \cite{googlenet}, are vulnerable to almost imperceptible intensity changes to the input images \cite{goodfellow6572explaining}. These small intensity changes can be regarded as adversarial attacks to CNNs. In medical image classification, the adversarial attacks can also fool CNN-based classifiers \cite{taghanaki2018vulnerability, finlayson2018adversarial}. Therefore, it is important to verify the robustness of CNNs before deploying them into practical use.

The verification of CNNs requires good understanding of the mechanism of adversarial attacks. In this paper, we aim at developing a novel method to generate adversarial examples which are able to attack CNN models for medical image segmentation. Generating adversarial examples to attack semantic image segmentation models is challenging because: 1) Semantic segmentation means assigning a label to each pixel (or voxel) instead of a single label per image as in conventional adversarial attacks typically described in computer vision scenarios. Therefore, attacking a segmentation model is more challenging than attacking a classification model; 2) It is not straightforward to evaluate the success of the attack. A good adversarial example for a classification model results in an incorrect prediction on the whole image while a good adversarial example for a segmentation model does not necessarily lead to an incorrect prediction for every pixel (voxel); 3) Conventional adversarial attacks perturb the image intensity by small amount, however, in medical imaging scenarios deformations are also useful to attack segmentation models. For instance, organs can be present in various configuration in images. Any segmentation model is therefore in principle susceptible to unseen poses or shapes of organs.

Generative adversarial networks (GAN) \cite{gan} and variational autoencoders (VAE) \cite{vae} are both unsupervised methods that can learn latent feature representations from training images. A GAN learns the latent feature representations implicitly while the VAE learns them explicitly. Training a GAN is difficult due to mode collapse and unreasonable results, e.g. a dog with two heads. In contrast, training a VAE is fairly simple. However, while a GAN can generate realistic images, images generated by a VAE are blurry because of the L2 loss employed during training. Inspired by these observations, we propose to combine the advantages of VAE and GAN to generate realistic and reasonable image deformations and appearance changes so that the transformed images can attack medical segmentation models.

Our main contributions can be summarised as follows: 1) We propose a novel approach to generate adversarial examples to attack the CNN model for abdominal organ segmentation in CT images; 2) We also measure the success of attack by means of observing significant reductions in the Dice score compared to ground truth segmentations; 3) The proposed approach attacking the segmentation model does not require any a-priori specification of particular attacks. In our application, we do not specify any organ which is attacked.

\section{Related Work}

The work in \cite{fischer2017adversarial}, \cite{metzen2017universal}, and \cite{Xie_2017_ICCV} represent state-of-the-art methods for attacking segmentation models. Fischer et al. \cite{fischer2017adversarial} proposed to attack segmentation models so that the models cannot segment object in a specified class (e.g. ignoring pedestrians on street). Metzen et al. \cite{metzen2017universal} proposed to generate adversarial examples so that the segmentation model incorrectly segments one cityscape as another one. The adversarial examples generated by these two methods attacked the segmentation models with specified targets, e.g. pedestrians. In contrast, Xie et al. \cite{Xie_2017_ICCV} proposed an approach to generate adversarial examples for image semantic segmentation and object detection without attacking targets. However, a random segmentation result should be specified so that the adversaries can be inferred. The adversarial attacks generated by these three methods often appear as pure noise that has no semantic meaning. Therefore, these attacks do not represent real-world situations that can occur in medical imaging applications.

\section{Our Approach}

\begin{figure}
\centering
\includegraphics[width=8cm]{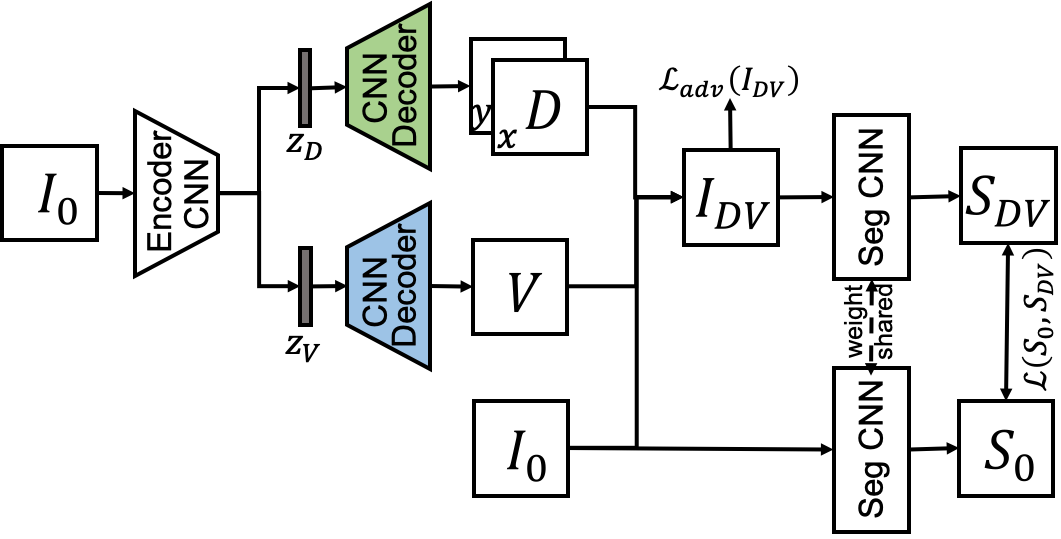}
\caption{Overview of the proposed framework. The CNN decoders in blue and green share the same architecture but they do not share weights.}\label{overview}
\end{figure}

We propose a novel end-to-end approach to generate adversarial examples for medical image segmentation scenarios. Formally, $\bm{I}_0$ is the original image ($H$ height and $W$ width) and $\bm{S}_0 \in \mathbb{R}^{H\times W \times (C+1)}$ is its segmentation given a fixed CNN-based segmentation model $f_{seg}(\cdot)$, i.e. $\bm{S}_0 = f_{seg}(\bm{I}_0)$. Here $C$ is the number of labels, e.g. the organs of interest. The adversarial attack model allows deformations $\bm{D}$ and intensity variations $\bm{V}$ applied to $\bm{I}_0$. $\bm{D}$ is a dense deformation field which is a displacement vector for each pixel (or voxel) while $\bm{V}$ is a smooth intensity perturbation which can be interpreted, e.g. as a bias field. Therefore, the transformed image after adversarial attack is given by
\begin{equation}
    \bm{I}_{DV} = \bm{I}_D + \bm{V} = f_D(\bm{I}_0,\bm{D}) + \bm{V}.
\end{equation}
Here $f_D(\cdot,\cdot)$ is the function which transforms $\bm{I}_0$ to $\bm{I}_D$ based on $\bm{D}$. Figure \ref{overview} shows the framework which learns appropriate $\bm{D}$ and $\bm{V}$ such that $\bm{I}_{DV}$ can attack the segmentation model. The whole framework consists of two key components: a CNN model for $\bm{I}_{DV}$ generation and it's learning algorithm.

\subsection{Model for generating adversarial attack}

The CNN architecture which generates the $\bm{I}_{DV}$ is similar to a multi-task VAE. First, $\bm{I}_0$ is processed by an encoding CNN resulting in several feature maps which are then used to learn a latent feature representation $\bm{z}_D \sim \mathcal{N}(\bm{\mu}_D,\bm{\sigma}_D^2)$ and $\bm{z}_V \sim \mathcal{N}(\bm{\mu}_V,\bm{\sigma}_V^2)$. $\bm{z}_D$ and $\bm{z}_V$ are then reconstructed to dense deformation field $\bm{D}$ and dense intensity variation $\bm{V}$ by two CNN decoders, respectively. The two decoding CNNs share the same architecture but they do not share weights. Learning $\bm{z}_D$ and $\bm{z}_V$ explicitly ensures the $\bm{I}_{DV}$ looks reasonable.

The dense deformation $\bm{D}$ consists of two channels of feature maps $\bm{D}_x$ and $\bm{D}_y$, representing pixel position changes in horizontal and vertical directions (i.e. $x$ and $y$ axis). In addition, we propose to limit the norm of $\bm{D}$ and $\bm{V}$ so that it is difficult to be perceived by human observers. To this end, the following to regularization terms are used:
\begin{equation}
\mathcal{L}(\bm{D}) =  \lambda_D\| \bm{D} \|_2^2; \quad \mathcal{L}(\bm{V}) = \lambda_V\| \bm{V} \|_2^2.
\end{equation}
$\lambda_D$ and $\lambda_V$ are two fixed hyper-parameters. The regularization ensures the smoothness of the deformation field $\bm{D}$ and intensity variation $\bm{V}$.

Each branch of the CNNs generating $\bm{I}_{DV}$ is different from a VAE since the input and output of the CNN are not the same. In fact, it is an image-to-image CNN and we can sample the learned latent space to generate multiple instances $\bm{D}$s and $\bm{V}$s. This idea is similar to the one proposed in \cite{kohl2018probabilistic} where a latent space was learned to sample multiple realistic image segmentations.

\subsection{Learning}
Since the ground truth of $\bm{D}$, $\bm{V}$, and $\bm{I}_{DV}$ are not available, it is not possible to learn the parameters of the encoding and decoding CNN in a explicit supervised manner. To address this problem, we propose to learn the parameters implicitly based on two conditions: First, we assume that $\bm{I}_{DV}$ should look realistic compared with $\bm{I}_0$. Secondly we assume that the accuracy of the segmentation $\bm{S}_{DV}$ should decreases significantly compared with $\bm{S}_0$. This decrease can be measured in terms of a reduction of Dice score.

An adversarial learning method is employed to ensure the $\bm{I}_{DV}$ looks realistic compared with $\bm{I}_0$. To this end, the $\bm{I}_{DV}$ generating CNN is regarded as a generator CNN, i.e. $\bm{I}_{DV} = f_{gen}(\bm{I}_0)$. An additional discriminator CNN $f_{disc}(\cdot)$ is used to predict the realism of $\bm{I}_{DV}$ compared with $\bm{I}_0$ \cite{jolicoeur2018relativistic}. Adversarial training $f_{gen}(\cdot)$ and $f_{disc}(\cdot)$ results in realistic $\bm{I}_{DV}$. We adopt the Wasserstein GAN (WGAN) \cite{wgan} loss function during the adversarial training. Formally,
\begin{equation}
\begin{aligned}
    \mathcal{L}_{adv}^{gen} & = f_{disc}(\bm{I}_0)-f_{disc}(\bm{I}_{DV}), \\
    \mathcal{L}_{adv}^{disc} & = f_{disc}(\bm{I}_{DV}) - f_{disc}(\bm{I}_0).
\end{aligned}
\end{equation}
The goal of this work is to generate $\bm{I}_{DV}$ which is able to attack a given segmentation CNN model, e.g. a U-Net \cite{unet}. This means that the segmentation results $\bm{S}_0$ and $\bm{S}_{DV}$ are different. Here $\bm{S}_{DV} = f_{seg}(\bm{I}_{DV})$. When training $f_{seg}(\cdot)$, we use cross-entropy as the loss function between $\bm{S}_0$ and the ground truth $\bm{S}_{GT}$, i.e. $f_{xent}(\bm{S}_0, \bm{S}_{GT})=-\sum_{c=0}^C\bm{S}_{GT}(c)\log(\bm{S}_0(c))$. This leads to satisfactory $\bm{S}_0$. To constrain the difference between $\bm{S}_0$ and $\bm{S}_{DV}$, we propose to use the following loss function:
\begin{equation}
    \mathcal{L}(\bm{S}_0, \bm{S}_{DV}) = (\xi - f_{xent}^M(\bm{S}_0, \bm{S}_{DV}))^2.
\end{equation}
Here $\xi$ is a hyper-parameter which controls the difference between $\bm{S}_0$ and $\bm{S}_{DV}$. If $\xi=0$, then $\bm{S}_{DV}$ tends to be similar to $\bm{S}_0$ so that $\bm{D}$ and $\bm{V}$ tend to be zero. In contrast, if $\xi$ is a very large number, then the norm of $\bm{D}$ and $\bm{V}$ are large that the discriminator CNN is difficult to fool. As such, the training process is likely to collapse. Therefore, $\xi$ should be within a proper range. In addition, we propose to mask the standard cross-entropy function so that the ROI of organs of interest is emphasized. Specifically, the masked cross-entropy function is
\begin{equation}
    f_{xent}^M(\bm{S}_{DV}, \bm{S}_0)=-\sum_{c=1}^CS_0(c)\ast\sum_{c=0}^C \bm{S}_0(c)\log(\bm{S}_{DV}(c)).
\end{equation}
Here, $M=\sum_{c=1}^CS_0(c)$ is the mask highlighting the organs of interest and $\ast$ is the element-wise product.

In summary, the loss functions of the whole framework are:
\begin{equation}
\begin{aligned}
    \mathcal{L}^{gen} & = \mathcal{L}_{adv}^{gen} + \mathcal{L}(\bm{S}_0, \bm{S}_{DV}) + \mathcal{L}(\bm{D}) + \mathcal{L}(\bm{V}), \\
    \mathcal{L}^{disc} & = \mathcal{L}_{adv}^{disc}.
\end{aligned}
\end{equation}

\subsection{Implementation Details}
In this paper, CNNs are implemented using Tensorflow. The adversarial learning is optimised using the RMSProp algorithm \cite{tieleman2012lecture}. The decay is 0.9 and $\epsilon=10^{-10}$. We use the fixed learning rate of $10^{-4}$ for both generator and discriminator CNNs. Batch normalization technique \cite{batchnorm} is used after convolutions. A leaky rectified linear unit (LReLU) is used as the nonlinear activation function to ease the adversarial training with $\alpha=0.2$. $\lambda_D$ and $\lambda_V$ are set as 0.1 and 0.01, respectively.

\section{Experiments and Results}

Experiments were performed on a abdominal CT dataset with multiple organs manually annotated by human experts. The image acquisition details and the involved patient demographics can be found in \cite{Tong2015}. The dataset consists of 150 subjects and for each subject the annotated organs include the pancreas, the kidneys, the liver, and the spleen. The dataset was randomly split into a training set, a validation set, and a testing set, which have 60, 15, and 75 subjects respectively. The voxel intensities of each subject were normalized to zero mean and unit standard deviation.

We trained a standard U-Net \cite{unet} to segment all abdominal organs. Due to limitations with GPU memory, the U-Net is based on 2D image, rather than 3D volumes. Following \cite{drinet}, the U-Net was trained on image patches and tested on image slices. The trained U-Net was used as the fixed CNN in this work to be subjected by adversarial attacks.

The Dice score was used to assess the segmentation quality for each organ. We define a 30\% decrease on the Dice score of an organ as a successful attack. Similar to \cite{nguyen2015deep,szegedy2013intriguing}, we compute the perceptibility $p$ of the adversarial perturbation $\bm{r} = \bm{I}_{DV}-\bm{I}_0$ by
\begin{equation}
    p=\frac{1}{HW}\sum_{i,j}|\bm{r}_{i,j}|.
\end{equation}
$p$ is the similarity of the real image to the synthetic image. The smaller the value of $p$ is, the less likely the adversarial perturbation is perceived by human observers.

\subsection{Adversarial Examples}

\begin{figure}
\centering
\includegraphics[width=10cm]{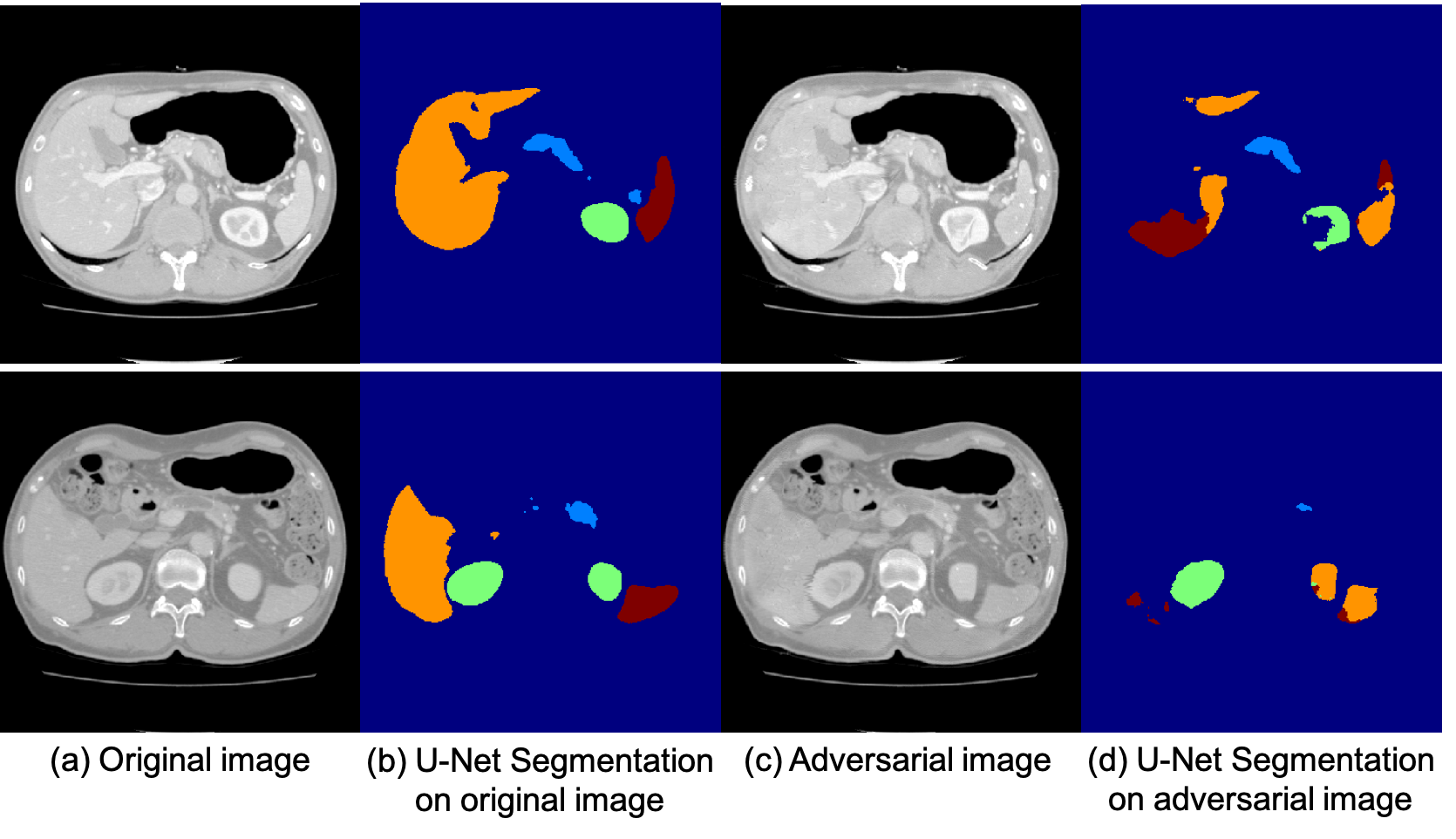}
\caption{Example CT abdominal image, the resulting U-Net segmentation, the generated adversarial CT images using the proposed method, and the resulting U-Net segmentation of the adversarial images. In the segmentations, the pancreas, the kidneys, the liver, and the spleen are depicted in blue, green, orange, and red, respectively. In this case, $\xi=2.0$.}\label{fig_results}
\end{figure}

\begin{figure}
\centering
\includegraphics[width=10cm]{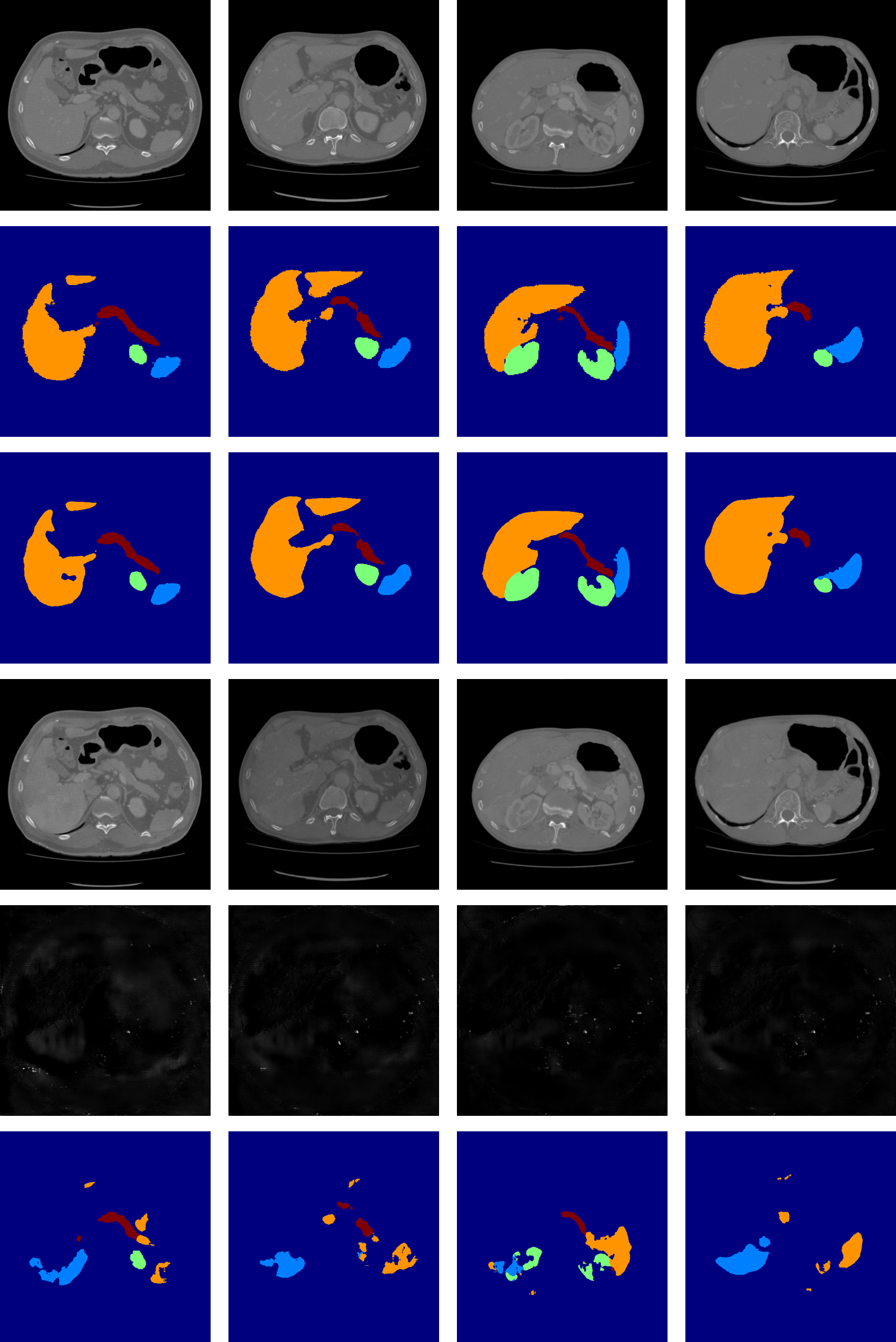}
\caption{Some additional visual results of the proposed method. Each column represents an individual case. $\bm{I}_0, \bm{S}_{GT}, \bm{S}_0, \bm{I}_{DV}, \bm{V}, \bm{S}_{DV}$ are shown from the top row to the bottom row.}
\label{fig_add_viz}
\end{figure}

By sampling the learned latent spaces, the deformation $\bm{D}$ and the intensity variation $\bm{V}$ are generated and therefore realistic adversarial examples are obtained. Figure \ref{fig_results} shows two such examples. Thanks to the regularization imposed on $\bm{D}$ and $\bm{V}$, both are smooth and difficult to recognise by humans. However, the derived adversarial examples attack the U-Net successfully. More examples are shown in Figure \ref{fig_add_viz}.

\subsection{Attacking the Segmentation Model}

Table \ref{tb_results} shows the segmentation results of the standard U-Net on multiple organs. The aforementioned success of the attacking model results in a 30\% decrease in terms of Dice score on every organ. We also listed this borderline of Dice score in Table \ref{tb_results}. In the proposed attack approach, $\xi$ is an important hyper-parameter deciding the success of attacking the U-Net. $\xi$ ranging from 0.5 to 3.0 was tested and the results are shown in Table \ref{tb_results}. The larger the $\xi$ is, the more the Dice scores decrease and the larger the perceptibility is. Using $\xi \ge 2.0$, the U-Net can be attacked successfully on all organs.

In terms of different organs, the segmentations on the pancreas and the kidneys are more difficult to be attacked compared to segmentations on the liver and the spleen. Specifically, the segmentations on the pancreas and the kidneys can be attacked when $\xi \ge 2.0$ while the segmentations on the liver and the spleen can be attacked when $\xi \ge 1.0$.

The proposed adversarial examples feature both deformations $\bm{D}$ and intensity variations $\bm{V}$. We studied the effect of $\bm{D}$ and $\bm{V}$ individually when $\xi=2.0$. The results are reported in Table \ref{tb_results}. For the kidneys, the deformation changes lead to more decrease of the Dice scores while on the other organs, the intensity variance has more impact on attacking the U-Net model. This means that the segmentation model is more sensitive to the intensity variance. The abdominal organs naturally vary in terms of pose on 2D image slices in the training set. Therefore, small deformations do not significantly decrease the Dice scores. In contrast, the intensity variations introduces shadows and artefacts which are likely to influence the segmentation CNN.

\begin{table}[ht!]
\caption{Abdominal organ segmentation comparison among different configurations in terms of the Dice score (\%).}
\label{tb_results}
\begin{center}
\begin{tabular}{ l | l  l  l  l | c }
\hline
\multirow{2}{*}{} & \multicolumn{4}{c|}{Dice} & \multirow{2}{*}{$p$}\\
                       & Pancreas & Kidneys & Liver & Spleen & \\
\hline
U-Net                  & 80.07 & 94.74 & 94.71 & 94.76 & -- \\
U-Net 30\% decrease    & 56.05 & 66.32 & 66.30 & 66.33 & -- \\
\hline
$\bm{I}_{DV}$ on U-Net ($\xi=0.5$) & 74.77 & 93.88 & 89.81 & 81.87 & 0.061 \\
$\bm{I}_{DV}$ on U-Net ($\xi=1.0$) & 70.66 & 90.06 & 59.51 & 37.12 & 0.060 \\
$\bm{I}_{DV}$ on U-Net ($\xi=1.5$) & 64.25 & 66.97 & 26.45 & 35.81 & 0.075 \\
$\bm{I}_{DV}$ on U-Net ($\xi=2.0$) & 53.59 & 60.07 & 11.19 & 17.21 & 0.074 \\
$\bm{I}_{DV}$ on U-Net ($\xi=2.5$) & 40.57 & 45.47 &  9.16 & 17.60 & 0.084 \\
$\bm{I}_{DV}$ on U-Net ($\xi=3.0$) & 31.49 & 43.43 &  5.82 & 26.58 & 0.085 \\
\hline
$\bm{I}_{D}$ on U-Net ($\xi=2.0$) & 70.46 & 82.04 & 75.06 & 70.47 & 0.056 \\
$\bm{I}_{V}$ on U-Net ($\xi=2.0$) & 68.15 & 88.25 & 50.05 & 46.94 & 0.061 \\
\hline
\end{tabular}
\end{center}
\end{table}

\section{Discussion and Conclusion}
In this paper, we have proposed a novel approach to generate adversarial examples to attack an existing CNN model for medical image segmentation. The generated adversarial examples include geometrical deformations to model anatomical variations as well as intensity variation which model appearance variations. These examples attack CNN-based segmentation models such as a U-Net \cite{unet} by decreasing the Dice score by a pre-specified amount. The training process is end-to-end without any predefined requirements. In fact, it can be replaced by any other CNN-based models. In the future, we will investigate the use of the  proposed approach to generate additional training images so that the segmentation model can be more robust and defend attacks. In addition, the proposed approach can be used to verify if an CNN model is robust or not. Specifically, our approach can generate adversarial examples for the CNN model. If the adversarial examples are reasonable and realistic, then the CNN model is not robust enough.
\bibliographystyle{splncs04}
\bibliography{adv_attack}

\begin{thebibliography}{10}
\providecommand{\url}[1]{\texttt{#1}}
\providecommand{\urlprefix}{URL }
\providecommand{\doi}[1]{https://doi.org/#1}

\bibitem{wgan}
Arjovsky, M., Chintala, S., Bottou, L.: Wasserstein generative adversarial
  networks. In: ICML. pp. 214--223 (2017)

\bibitem{drinet}
Chen, L., Bentley, P., Mori, K., Misawa, K., Fujiwara, M., Rueckert, D.:
  {DRINet for medical image segmentation}. IEEE TMI  \textbf{37}(11),
  2453--2462 (2018)

\bibitem{finlayson2018adversarial}
Finlayson, S.G., Kohane, I.S., Beam, A.L.: Adversarial attacks against medical
  deep learning systems. arXiv preprint arXiv:1804.05296  (2018)

\bibitem{fischer2017adversarial}
Fischer, V., Kumar, M.C., Metzen, J.H., Brox, T.: Adversarial examples for
  semantic image segmentation. arXiv preprint arXiv:1703.01101  (2017)

\bibitem{gan}
Goodfellow, I., Pouget-Abadie, J., Mirza, M., Xu, B., Warde-Farley, D., Ozair,
  S., Courville, A., Bengio, Y.: Generative adversarial nets. In: NIPS. pp.
  2672--2680 (2014)

\bibitem{goodfellow6572explaining}
Goodfellow, I.J., Shlens, J., Szegedy, C.: Explaining and harnessing
  adversarial examples. In: ICLR (2015)

\bibitem{batchnorm}
Ioffe, S., Szegedy, C.: {Batch normalization: Accelerating deep network
  training by reducing internal covariate shift}. In: ICML. pp. 448--456 (2015)

\bibitem{jolicoeur2018relativistic}
Jolicoeur-Martineau, A.: {The relativistic discriminator: a key element missing
  from standard GAN}. arXiv preprint arXiv:1807.00734  (2018)

\bibitem{vae}
Kingma, D.P., Welling, M.: Auto-encoding variational bayes. arXiv preprint
  arXiv:1312.6114  (2013)

\bibitem{kohl2018probabilistic}
Kohl, S., Romera-Paredes, B., Meyer, C., De~Fauw, J., Ledsam, J.R., Maier-Hein,
  K., Eslami, S.A., Rezende, D.J., Ronneberger, O.: {A probabilistic U-Net for
  segmentation of ambiguous images}. In: NIPS. pp. 6965--6975 (2018)

\bibitem{metzen2017universal}
Metzen, J.H., Kumar, M.C., Brox, T., Fischer, V.: Universal adversarial
  perturbations against semantic image segmentation. In: ICCV. pp. 2755--2764
  (2017)

\bibitem{nguyen2015deep}
Nguyen, A., Yosinski, J., Clune, J.: Deep neural networks are easily fooled:
  High confidence predictions for unrecognizable images. In: CVPR. pp. 427--436
  (2015)

\bibitem{unet}
Ronneberger, O., Fischer, P., Brox, T.: U-net: convolutional networks for
  biomedical image segmentation. In: MICCAI. pp. 234--241 (2015)

\bibitem{googlenet}
Szegedy, C., Liu, W., Jia, Y., Sermanet, P., Reed, S., Anguelov, D., Erhan, D.,
  Vanhoucke, V., Rabinovich, A.: Going deeper with convolutions. In: CVPR.
  pp.~1--9 (2015)

\bibitem{szegedy2013intriguing}
Szegedy, C., Zaremba, W., Sutskever, I., Bruna, J., Erhan, D., Goodfellow, I.,
  Fergus, R.: Intriguing properties of neural networks. In: ICLR (2014)

\bibitem{taghanaki2018vulnerability}
Taghanaki, S.A., Das, A., Hamarneh, G.: {Vulnerability analysis of chest X-ray
  image classification against adversarial attacks}. In: Understanding and
  Interpreting Machine Learning in Medical Image Computing Applications, pp.
  87--94. Springer (2018)

\bibitem{tieleman2012lecture}
Tieleman, T., Hinton, G.: Lecture 6.5-rmsprop: Divide the gradient by a running
  average of its recent magnitude. COURSERA: Neural networks for machine
  learning  \textbf{4}(2),  26--31 (2012)

\bibitem{Tong2015}
Tong, T., Wolz, R., Wang, Z., Gao, Q., Misawa, K., Fujiwara, M., Mori, K.,
  Hajnal, J.V., Rueckert, D.: Discriminative dictionary learning for abdominal
  multi-organ segmentation. Medical Image Analysis  \textbf{23}(1),  92--104
  (2015)

\bibitem{Xie_2017_ICCV}
Xie, C., Wang, J., Zhang, Z., Zhou, Y., Xie, L., Yuille, A.: Adversarial
  examples for semantic segmentation and object detection. In: ICCV. pp.
  1369--1378 (2017)

\end{thebibliography}

\end{document}